\documentclass[11pt,twoside,a4paper,cmspaper,final,collab]{cms-tdr}
\def\svnVersion{4473969}\def\svnDate{2026/03/05}\def\cmsCernNoTag{CERN-EP-2025-213}\def\cmsCernDate{\today}\def\cmsMessage{Submitted to Physical Review Letters}
\begin{document}\cmsNoteHeader{EXO-24-007}

\newcommand{\PQl}{{\HepParticle{q_{l}}{}{}}\xspace}
\newcommand{\PAQl}{{\HepAntiParticle{q}{l}{}}\xspace}
\newcommand{\PA}{{\HepParticle{A}{}{}}\xspace}
\newcommand{\PAprime}{\ensuremath{\PA^{\prime}}\xspace}

\newcommand{\PM}{{\HepParticle{M}{}{}}\xspace}
\newcommand{\PDM}{{\HepParticle{DM}{}{}}\xspace}
\newcommand{\PALP}{{\HepParticle{ALP}{}{}}\xspace}

\newcommand{\mPALP}{\ensuremath{m_{\PALP}}\xspace}
\newcommand{\mPAprime}{\ensuremath{m_{\PAprime}}\xspace}
\newcommand{\mPDM}{\ensuremath{m_{\PDM}}\xspace}
\newcommand{\mPH}{\ensuremath{m_{\PH}}\xspace}
\newcommand{\mphi}{\ensuremath{m_{\phi}}\xspace}
\newcommand{\mPM}{\ensuremath{m_{\PM}}\xspace}
\newcommand{\mPZ}{\ensuremath{m_{\PZ}}\xspace}
\newcommand{\mPZpr}{\ensuremath{m_{\PZpr}}\xspace}

\newcommand{\fxfx}{{\textsc{FxFx}}\xspace}
\newcommand{\MADSPIN}{\textsc{MadSpin}\xspace}
\newcommand{\HJMINLO}{{\textsc{HJ-MiNLO}}\xspace}

\newcommand{\mSD}{\ensuremath{m_{\mathrm{SD}}}\xspace}
\newcommand{\abseta}{\ensuremath{\abs{\eta}}}

\newcommand{\gq}{\ensuremath{g_{\PQq}}\xspace}
\newcommand{\gqphi}{\ensuremath{g_{\PQq\phi}}\xspace}
\newcommand{\gqA}{\ensuremath{g_{\PQq\PA}}\xspace}
\newcommand{\gb}{\ensuremath{g_{\PQb}}\xspace}

\newcommand{\PNET}{{\textsc{ParticleNet}}\xspace}
\newcommand{\PNscore}[1]{\ensuremath{P_{#1}}}
\newcommand{\PNbb}{\PNscore{\PQb\PAQb}}
\newcommand{\PNcc}{\PNscore{\PQc\PAQc}}
\newcommand{\PNll}{\PNscore{\PQl\PAQl}}
\newcommand{\PNqq}{\PNscore{\PQl\PAQl}}
\newcommand{\PNqcd}{\PNscore{\text{QCD}}}

\newcommand{\pxx}{\ensuremath{p_{\text{2-prong}}}{}\xspace}
\newcommand{\pbvcl}{\ensuremath{p_{\PQb\PAQb}}{}\xspace}

\newcommand{\TFMC}{\ensuremath{T_\text{MC}}\xspace}
\newcommand{\TFres}{\ensuremath{T_\text{res}}\xspace}

\newcommand{\thetaW}{\ensuremath{\theta_{\PW}}\xspace}

\ifthenelse{\boolean{cms@external}}{
\newcommand{\cmsLeft}{upper\xspace}
\newcommand{\cmsRight}{lower\xspace}
\renewcommand{\appendixname}{End Matter}
}{
\newcommand{\cmsLeft}{left\xspace}
\newcommand{\cmsRight}{right\xspace}
}
\newcommand{\cmsEndMatter}{\appendixname~\ref{appendix}}

\cmsNoteHeader{EXO-24-007}

\title{Search for low-mass vector and scalar resonances decaying into a quark-antiquark pair in proton-proton collisions at \texorpdfstring{$\sqrt{s}=13\TeV$}{sqrt(s)=13 TeV}}

\date{\today}

\abstract{
A search for resonances with masses from 50 to 300\GeV decaying into a quark-antiquark pair is presented. The search uses proton-proton collision data at $\sqrt{s}=13\TeV$ collected by the CMS experiment at the CERN LHC in 2016--2018, corresponding to an integrated luminosity of 138\fbinv. Two coupling scenarios are considered, with the resonances coupled either equally to all flavors of quarks or preferentially to bottom quarks. The search targets resonances produced in association with hard initial-state radiation, resulting in a large-radius jet with a two-pronged substructure. The \PNET algorithm is used to distinguish resonance decays to bottom quark pairs from lighter quark pairs and to suppress background processes. The invariant jet mass spectrum is scrutinized for peaking excesses over a falling background. No evidence for such resonances is observed. Limits are set on the couplings of new scalar and vector resonances to quarks, representing the most stringent limits to date in the mass range of 50--250\GeV.
}
\hypersetup{
pdfauthor={CMS Collaboration},%
pdftitle={Search for low-mass vector and scalar resonances decaying into a quark-antiquark pair in proton-proton collisions at sqrt(s)=13 TeV},%
pdfsubject={CMS},%
pdfkeywords={CMS, dijet, boosted, ParticleNet}}

\maketitle 

New resonances coupled to a pair of quarks are a common signature of theories of beyond the standard model (SM) physics. 
Additional \PWpr or \PZpr bosons are proposed in models with an extended electroweak (EW) gauge sector, such as left-right symmetric SM extensions~\cite{Mohapatra:1981wj,Eichten:1984eu}. 
Randall--Sundrum models with a warped extra dimension predict Kaluza--Klein excitations of gravitons that couple to quark-antiquark pairs~\cite{ref_rsg}. 
Simplified models of weakly interacting dark matter propose new mediator particles that connect the SM to a dark sector, which could couple preferentially to quarks~\cite{Abercrombie:2015wmb,dmwg_dmsimp}. 
Such particles could be resonantly produced with large cross sections at hadron colliders. 
Consequently, searches for new dijet resonances have been pursued extensively~\cite{ATLAS:2024kpy,pdg2024,CMS:2024zqs}. 

This Letter presents a search for new, narrow dijet resonances with masses in the range 50--300\GeV, using proton-proton ($\Pp\Pp$) collision data at $\sqrt{s}=13\TeV$ collected by the CMS experiment at the CERN LHC in 2016--2018.
The data set corresponds to an integrated luminosity of 138\fbinv.
Two simplified benchmark scenarios are considered: a spin-1 boson, \PZpr, coupled equally to all quark flavors with a coupling constant \gq, and a spin-0 boson---either scalar, $\phi$, or pseudoscalar, \PA---with quark couplings equal to the SM Yukawa couplings times a flavor-universal scaling factor, \gqphi or \gqA, respectively~\cite{Abercrombie:2015wmb}. 
In the spin-0 scenario, which is motivated by minimal flavor violation~\cite{mfv}, the resonance is produced via gluon fusion and decays dominantly to bottom quark-antiquark pairs ($\PQb\PAQb$). 

The primary background consists of SM events containing two or more jets produced through the strong interaction, referred to as quantum chromodynamics (QCD) multijet events. 
To circumvent experimental readout bandwidth limitations imposed by the large QCD background rate, this analysis searches for resonances produced with transverse momentum (\pt) greater than 500\GeV as a result of hard initial-state radiation (ISR). 
The ISR consists dominantly of quark or gluon radiation, but no specific ISR requirements are imposed.
Because of the large \pt, the resonance decay products are collimated in the detector, and are reconstructed as a single large-radius jet with a two-pronged substructure. 
The signature of the resonance is a peaking excess in the large-radius jet invariant mass spectrum.
For the coupling values considered in this analysis, the resonance width is negligible compared to the experimental resolution.
The \PNET (PN) algorithm~\cite{particlenet} is used to distinguish the resonances from the background of jets initiated by light quarks or gluons, and significantly enhances the sensitivity compared to analytic variables, such as the energy correlation functions~\cite{ecf} used in previous searches~\cite{CMS-boosted5}. 
The PN algorithm is also used to define separate signal regions (SRs) targeting resonances decaying to \bbbar pairs and to charm or light (\ie, up, down, and strange) quark pairs ($\PQq\PAQq)$.

The CMS apparatus~\cite{CMS:2008xjf,CMS:2023gfb} is a multipurpose, nearly hermetic detector, designed to trigger on~\cite{CMS:2020cmk,CMS:2016ngn,CMS:2024aqx} and identify electrons, muons, photons, and hadrons~\cite{CMS:2020uim,CMS:2018rym,CMS:2014pgm}. The ``particle-flow'' algorithm~\cite{CMS:2017yfk} aims to reconstruct all individual particles in an event, 
combining information provided by the all-silicon inner tracker and by the crystal electromagnetic and brass-scintillator hadron calorimeters, 
operating inside a 3.8\unit{T} superconducting solenoid, 
with data from the gas-ionization muon detectors embedded in the flux-return yoke outside the solenoid. 
The reconstructed particles are used to build \PGt leptons, jets, and missing transverse momentum (\ptvecmiss)~\cite{CMS:2018jrd,CMS:2016lmd,CMS:2019ctu}.
The data are analyzed in four separate data-taking periods corresponding to significant changes in the detector and LHC configurations. 
Tabulated results are provided in the HEPData record for this analysis~\cite{hepdata}.

The signal and background processes are simulated at the matrix element level using several Monte Carlo (MC) event generators. 
Matrix-element level events are showered using \PYTHIA~8.230~\cite{Sjostrand:2014zea} with the CP5 tune~\cite{cp5}. 
For events generated at leading order (LO), jets from the matrix element calculations are matched to the parton shower using the MLM prescription~\cite{mlm}, while events generated at next-to-LO (NLO) in QCD use the \fxfx merging procedure~\cite{fxfx}. 
All samples use the NNPDF3.1 parton distribution function (PDF) set at next-to-NLO accuracy. 
Additional $\Pp\Pp$ interactions within the same or adjacent bunch crossings (pileup) are simulated using \PYTHIA, with the multiplicity distribution matching that in data. 
The response of the CMS detector is modeled using \GEANTfour~\cite{geant}.
Separate simulation samples are produced for each data-taking period. 

The $\PW(\qqbar)$+jets ($\PZ(\qqbar)$+jets) process is modeled at LO using \MGvATNLO~2.6.5, with up to three (four) extra partons in the matrix element. 
The cross sections are corrected to higher order in the perturbative expansion as a function of the generator-level boson \pt: the QCD NLO corrections are derived using \MGvATNLO, with up to two extra partons in the matrix element calculations, and the EW NLO corrections are taken from theoretical calculations in Refs.~\cite{kallweit2015nlo,Kallweit_2016,kallweit2015nlo2,Lindert_2017}.
The top quark-antiquark (\ttbar) and single top quark processes are modeled at NLO in QCD using \POWHEG~2.0~\cite{Frixione:2007nw,Frixione:2007vw,Alioli:2010xd,Frederix:2012dh,Re:2010bp}. 
The QCD multijet process is modeled at LO using \MGvATNLO with up to four partons at the matrix element level. 
The diboson processes are simulated using \PYTHIA, with cross sections corrected to NLO accuracy in QCD using \MCFM~7.0~\cite{Campbell:2010ff}.
The contribution from SM Higgs boson (\PH) production via gluon-gluon fusion is modeled using the \HJMINLO generator, including finite top quark mass effects and with $\mPH=125\GeV$~\cite{Neumann:2018bsx,Monni:2019whf,Monni:2020nks,becker2021precise}. 
The vector boson fusion and $\ttbar\PH$ production modes are simulated with \POWHEG~\cite{Nason:2009ai,Luisoni:2013kna,Hartanto:2015uka} at NLO in QCD. 
Small contributions from $\PW(\Pl\PGnl)$ and the Drell--Yan process are simulated at LO using \MGvATNLO.
Finally, the signal processes are modeled at LO using \MGvATNLO. 
Vector \PZpr boson production via quark-antiquark annihilation is simulated with up to two extra partons at the matrix element level, while the scalar and pseudoscalar production via gluon fusion is simulated with up to one extra parton. 
The \PZpr boson cross sections are multiplied by a factor of 1.2 to account for NLO contributions, calculated using \MGvATNLO.

Small- and large-radius hadronic jets are clustered from reconstructed particles using the anti-\kt algorithm~\cite{Cacciari:2008gp,Cacciari:2011ma} with distance parameters of 0.4 (AK4 jets) and 0.8 (AK8 jets), respectively.
The signal candidates correspond to AK8 jets. 
Jet momentum is determined as the vectorial sum of all particle momenta in the jet, and is found from simulation to be, on average, within 5 to 10\% of the true momentum over the entire \pt spectrum and detector acceptance. 
The effects of pileup are mitigated using the charged hadron subtraction and pileup-per-particle identification algorithms for the AK4 and AK8 jets, respectively~\cite{Bertolini:2014bba,Sirunyan:2020foa}.
Corrections are applied to the jet energy to account for the detector response~\cite{CMS:2016lmd}.

The soft-drop jet mass~\cite{softdrop}, \mSD, and the mass-decorrelated PN jet tagging algorithm~\cite{particlenet} are used to identify the signal resonances and distinguish them from the dominant QCD background. 
The soft-drop mass algorithm, with parameters $z_{\text{cut}}=0.1$ and $\beta=0$, reduces the mass of quark and gluon jets while preserving the mass of jets due to heavy resonances decaying to quark-antiquark pairs, and provides a mass resolution of approximately 9--14\%. 
The PN algorithm is used to distinguish signal jets from the multijet background and identify their flavor. 
The algorithm is based on a graph convolutional neural network with the jet constituents and secondary vertices as inputs.
The network is trained on simulation, with the training samples tailored to reduce the correlation between the tagger performance and the jet mass in QCD events. 
The outputs comprise four nodes, $\PNscore{i}$, where $i$ corresponds to the four processes considered: the QCD background, and a boosted resonance, \PX, decaying to
bottom, charm, and light quark-antiquark pairs, which are denoted $\PX\to\PQb\PAQb$, $\PX\to\PQc\PAQc$, and $\PX\to\PQl\PAQl$, respectively.
The outputs are combined to form two discriminants: 
\begin{align}
    \pxx &= \frac{\PNbb+\PNcc+\PNqq}{\PNbb+\PNcc+\PNqq+\PNqcd}\quad \text{and} \\
    \pbvcl &= \frac{\PNbb}{\PNbb + \PNcc + \PNqq}.
\end{align}
The \pxx discriminant distinguishes two-pronged resonances of any flavor from the QCD background, while the \pbvcl discriminant distinguishes \bbbar resonances from $\PQc\PAQc$ and $\PQl\PAQl$ resonances. 

Events are selected online using a set of triggers imposing minimum thresholds on jet activity in the event.
The primary trigger requires an AK8 jet with a trimmed mass~\cite{jetTrimming} greater than 30\GeV and $\pt>400\GeV$.
The trigger requirements are fully efficient with respect to the offline selection for AK8 jets with $\pt>500\GeV$. 
The offline selection requires at least one AK8 jet with $\pt>500\GeV$, $\mSD>30\GeV$, and $\abseta<2.5$.
If an event contains more than one such jet, the jet with the highest \pxx is chosen as the signal candidate.

Additional selection criteria are used to suppress the \ttbar background.
The background due to \ttbar events where a \PW boson decays leptonically is reduced by vetoing events containing a charged lepton passing one of the following criteria. 
Electrons (muons) must have $\pt>10\GeV$, $\abseta<2.5$ (2.1), and pass the loose identification criteria described in Ref.~\cite{CMS:2020uim} (\cite{CMS:2018rym}).
The electrons and muons are also required to be isolated from other particles in the events: the isolation variable, defined for electrons (muons) as the pileup-corrected \pt sum of charged hadrons and neutral particles within a cone in the $\eta$-$\phi$ plane of $\Delta R = 0.3$ (0.4) around the lepton divided by the lepton \pt, must be less than 0.15 (0.25). 
Hadronically decaying tau leptons, reconstructed using the hadron-plus-strips algorithm~\cite{CMS:2018jrd}, must pass the tight working point of the \textsc{DeepTau} algorithm~\cite{cms_deeptau} and have $\pt>20\GeV$ and $\abseta<2.3$. 
Events are also vetoed if they contain six or more AK4 jets with $\pt>30\GeV$ and $\abseta<5$. 
Finally, events must not contain an AK4 jet with $\pt>30\GeV$ and $\abseta<2.5$ that passes the medium working point of the \textsc{DeepJet} \PQb tagging algorithm~\cite{cms_deepjet2,CMS-DP-2023-005} and satisfies $\abs{\Delta\phi}>2\pi/3$ with respect to the selected AK8 jet.

The events are categorized into three regions based on the PN discriminants. 
First, events are divided based on their \pxx score, with a working point corresponding to a mistagging rate of 1\% for QCD events. 
Events failing the \pxx requirement are used for the QCD multijet background estimation, and constitute the QCD control region (CR). 
The events that pass the \pxx requirement are further divided using the \pbvcl score.
The high-\pbvcl SR targets the $\phi/\PA\to\bbbar$ and $\PZpr\to\bbbar$ signatures, while the low-\pbvcl SR targets \PZpr boson decays to light and charm quarks. 
For $\PZpr\to\PQb\PAQb$, $\PZpr\to\PQc\PAQc$, and $\PZpr\to\PQl\PAQl$ events with masses between 50 and 300\GeV, the efficiencies of the \pxx requirement are in the ranges 72--38\%, 52--19\%, and 24--5\%, respectively. 
For signal events passing the \pxx requirement, more than 95\% of events are assigned to the correct high- or low-\pbvcl SR. 

Finally, the events are divided into 62 bins in \mSD from 40 to 350\GeV, and five \pt bins with boundaries of 500, 550, 600, 700, 800, and 1200\GeV.
Only bins with the barycenter satisfying $-5.5>\rho>-2.1$ are considered, where $\rho=2\ln(\mSD/\pt)$, yielding 203 total bins.
The lower $\rho$ threshold avoids regions where generator predictions become less reliable because of nonperturbative effects~\cite{ATLASrho}, while the upper threshold avoid regions where the resonance decay products are close to the jet boundary.

The dominant background in all SRs is QCD multijet production. 
The $\PW/\PZ(\qqbar)$+jets processes contribute significant resonant backgrounds.
The \ttbar process contributes a small nonresonant background primarily in the high-\pbvcl SR. 
Additional minor backgrounds arise from \PH, single top quark, diboson, and Drell--Yan processes.
The QCD background is estimated using control samples in data, while all other backgrounds are estimated using MC simulation.

The cross sections of the SM backgrounds from the \PW, \PZ, and \PH boson processes are corrected to higher orders, as described above.
The simulated efficiency of the 2-pronged selection criterion, \mSD scale, and \mSD resolution are corrected to match data using a CR targeting boosted, hadronically decaying \PW bosons in \ttbar events where one \PW boson decays hadronically and the other leptonically. 
Events are required to have one muon with $\pt>53\GeV$ passing the tight identification requirements described in Ref.~\cite{CMS:2018rym} and a \PQb-tagged jet in the opposite hemisphere from the selected AK8 jet, as well as $\abs{\ptvecmiss}>40\GeV$. 
Additionally, the vector sum of the muon \pt and \ptvecmiss must have a magnitude of at least 200\GeV.
The corrections are extracted from a template fit to the \mSD distributions in events passing and failing the 2-pronged requirement. 
The templates corresponding to the \mSD scale and resolution variations are constructed using linear morphing~\cite{morphing}. 
The measured efficiency correction ranges from 0.72 to 0.91, depending on the data-taking period, with the uncertainties of 0.07--0.14.
The \mSD scale ranges $-$0.5--0.5\%, with the uncertainties of 0.5--0.9\%. 
The \mSD resolution scale factor, defined as a multiplicative correction to the reconstructed width of the resonances, ranges from 1.04 to 1.22, with the uncertainties of 5--10\%. 
Additional corrections are applied to the modeling of the efficiency of the \pbvcl requirement for \bbbar resonances, including the signal processes, $\PZ\to\bbbar$, and $\PH\to\bbbar$, as well as the mistagging rate for resonances decaying to a light quark-antiquark pair. 
Based on a CR targeting gluon splitting to $\bbbar$~\cite{CMS-PAS-BTV-22-001}, a conservative 30\% uncertainty is assigned. 
These uncertainties are constrained in data during the signal extraction by the SM \PW and \PZ resonances. 

The SM background from \ttbar production is taken from simulation. 
The overall normalization and the efficiencies of the \pxx and \pbvcl requirements are corrected using a \ttbar CR targeting events where one \PW boson decays leptonically and the other hadronically, and the decay products of the hadronically decaying \PQt quark are contained within the selected AK8 jet. 
The events are categorized into the same three PN categories as the SR, which are fitted simultaneously with the SRs and the QCD CR during the signal extraction. 
The corrections are implemented as two nuisance parameters, one modifying the mistagging rate of \PQt quark jets and the other scaling the overall normalization of the \ttbar and single top quark backgrounds; both are consistent with unity in the fit.

The QCD multijet background is estimated from data using the QCD CR, similarly to the method described in Ref.~\cite{CMS:2024ddc}. 
For each SR, the shape of the QCD background in \pt and $\rho$ is taken from the QCD CR, modified by two polynomial transfer functions to account for small shape differences between the regions. 
For a given bin $i$ with the barycenter $(\pt^i,\,\rho^i)$, the QCD background prediction is given by:
\begin{equation}
N_{\text{SR}}^i =  N_{\text{CR}}^{\text{data},i} \; R_{\text{SR/CR}}^{\text{MC}} \; \TFMC(\pt^i,\rho^i) \; \TFres(\pt^i,\rho^i),
\end{equation}
where $N_{\text{CR}}^{\text{data},i}$ is a free fit parameter corresponding to the expected QCD background in bin $i$ of the QCD CR and $R_{\text{SR}\text{/CR}}^{\text{MC}}$ is the overall ratio of SR to QCD CR events in simulation. 
The first transfer function, $\TFMC(\pt^i,\rho^i)$, accounts for 
residual dependence of the tagging efficiency on \pt and $\rho$ in simulation, and is derived from QCD simulation. The coefficients are constrained by their fitted uncertainties during the signal extraction.
The second transfer function, $\TFres(\pt^i,\rho^i)$, accounts for residual differences in the tagging efficiency between data and simulation, with coefficients treated as unconstrained parameters during the signal extraction.
Both transfer functions are parameterized as Bernstein polynomials, with the orders determined using a Fisher $F$-test~\cite{ftest}, separately for each SR and data-taking period.
For \TFres (\TFMC), the orders range from 0 to 2 for \pt and 0 to 3 (0 to 4) for $\rho$.

Systematic uncertainties account for possible mismodeling of the background and signal processes and are included as nuisance parameters in the signal extraction procedure. 
The dominant systematic uncertainty arises from the modeling of the QCD multijet background. 
The uncertainty arises primarily from the unconstrained coefficients of \TFres and, to a lesser extent, from \TFMC.

A set of systematic uncertainties relate to the modeling of resonant processes, including the signal processes and SM \PW, \PZ, and \PH production.
The uncertainty in the 2-pronged tagging efficiency is incorporated in the fit as a constrained nuisance parameter with a nominal value of unity and an uncertainty equal to the deviation of the measured correction from unity, ranging from 10 to 28\% depending on the data-taking period. 
Similarly, the efficiency and mistagging rate of the \pbvcl requirements are modified by nuisance parameters with a nominal value of unity and a conservative uncertainty of 30\%. 
The uncertainties in the \mSD scale and resolution range up to 0.9\% and 12\%, respectively. 
Uncertainties are also assigned to the theoretical cross sections of the \PW and \PZ processes to account for missing higher-order corrections and mixed QCD-EW effects, according to the prescription described in Ref.~\cite{Lindert_2017}. 
These uncertainties are constrained in data by the SM \PW and \PZ resonances.

Additional small experimental systematic uncertainties are included, amounting to less than a few percent.
The uncertainty in the integrated luminosity is partially correlated across the data-taking periods, with individual uncertainties of 1.2--2.5\% and an overall uncertainty of 1.6\%~\cite{CMS-LUM-17-003,CMS-PAS-LUM-17-004,CMS-PAS-LUM-18-002}.
The jet energy scale and resolution uncertainties, which affect the jet \pt, range from 0.5 to 4.0\% and are treated as independent from the \mSD uncertainties.
Uncertainties due to the modeling of lepton identification and isolation criteria, pileup, and the \PQb tagging of AK4 jets, as well as the limited number of simulated events, are included, but have a negligible impact on the signal extraction.

\begin{figure}[htbp]
    \centering
    \includegraphics[width=0.48\textwidth]{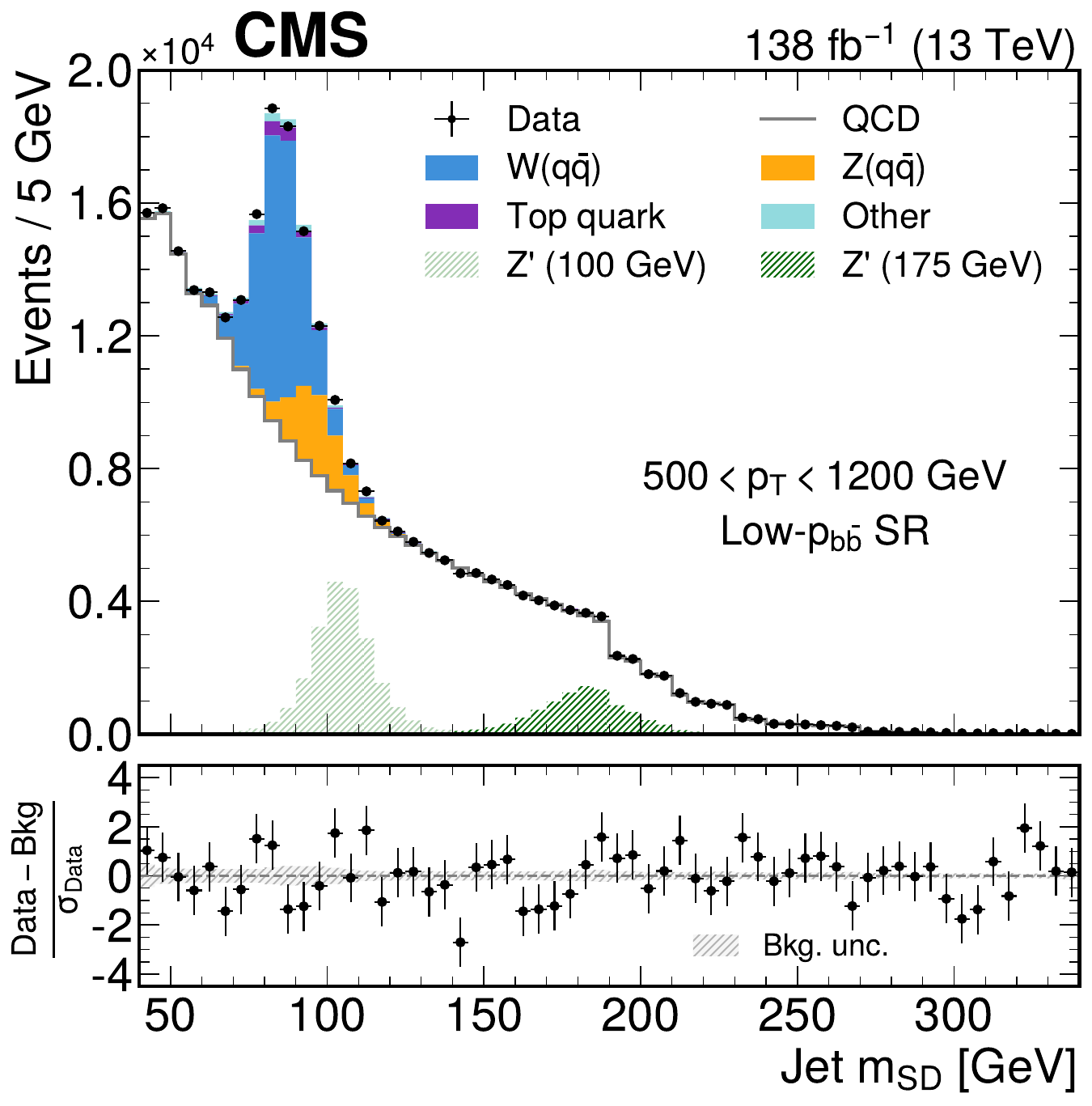}
    \includegraphics[width=0.48\textwidth]{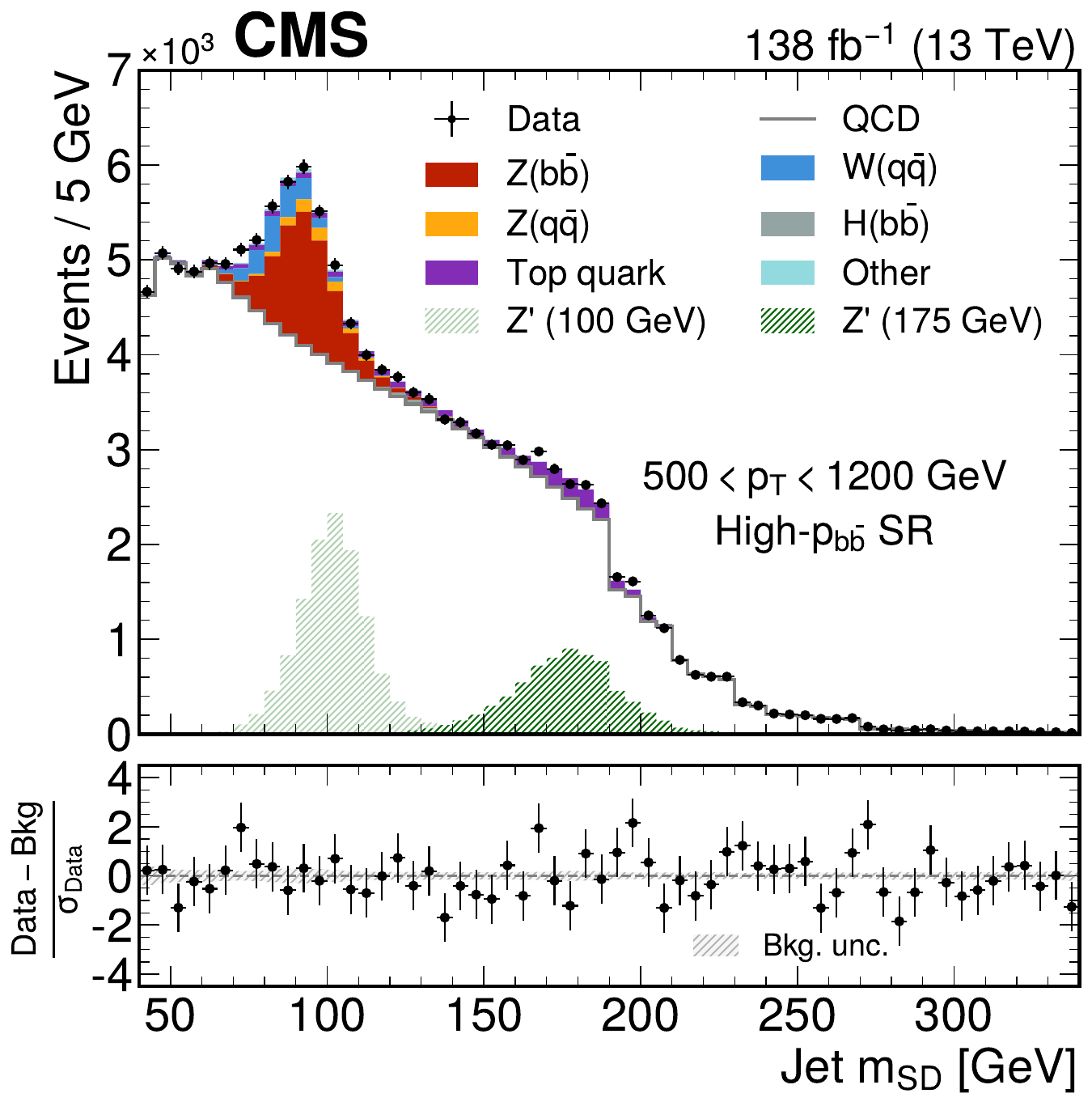}

    \caption{
        The jet \mSD distributions in the low-\pbvcl (\cmsLeft) and high-\pbvcl (\cmsRight) SRs, fitted with the background-only hypothesis. 
        The distributions are summed across all five \pt bins and the four data-taking periods.
        The steplike features are due to the $\rho$ requirement, which results in a different \mSD range for each \pt bin.
        The hatched area shows the expected \PZpr signal yields for $\gq=0.25$. 
        The lower panel shows the residual difference between the data and the overall background, divided by the statistical uncertainty of the data; the gray band represents the systematic uncertainty in the background estimation. 
    }
    \label{fig:mSD}
\end{figure}

For each signal hypothesis, a binned maximum likelihood fit of the signal and background templates is performed to the \mSD distributions, simultaneously in the five \pt bins of the high-\pbvcl SR, the low-\pbvcl SR, and the QCD CR, plus the three categories of the \ttbar CR. 
The statistical analysis uses the CMS \textsc{Combine} tool~\cite{CMS:2024onh}.
In total, there are 72 \mSD distributions, 18 per data-taking period, that are simultaneously fit.
The signal templates are constructed from MC simulation. 
The background-only fit is shown in Fig.~\ref{fig:mSD}, which displays the \mSD distributions in the low- and high-\pbvcl SRs, summing the four data-taking periods and the five \pt bins.
The SM \PW and \PZ bosons are clearly observed in the SRs. 
The high-\pbvcl SR also has a contribution from the \ttbar background, which is negligible in the low-\pbvcl SR. 
The goodness of fit is determined by fitting pseudoexperiment data sets drawn with posterior nuisance parameters. 
Using a saturated goodness of fit test statistic~\cite{Cousins:2018tiz}, the $p$-value of the background-only model hypothesis is 0.194. 

No significant excess is observed above the background predictions.
For the \PZpr model with decays to all quark flavors, the most significant deviations from the background expectations correspond to local (global) significances of $2.8\sigma$ ($1.7\sigma$) at $\mPZpr=75\GeV$ and $2.8\sigma$ ($1.6\sigma$) at $\mPZpr=225\GeV$. 
For the spin-0 model with decays only to \bbbar pairs, the most significant deviation corresponds to a local (global) significance of $2.6\sigma$ ($1.4\sigma$) at $\mphi=70\GeV$.
The global significance values are determined using a \mSD resolution of 10\%. 

\begin{figure}[htbp]
    \centering
    \includegraphics[width=0.48\textwidth]{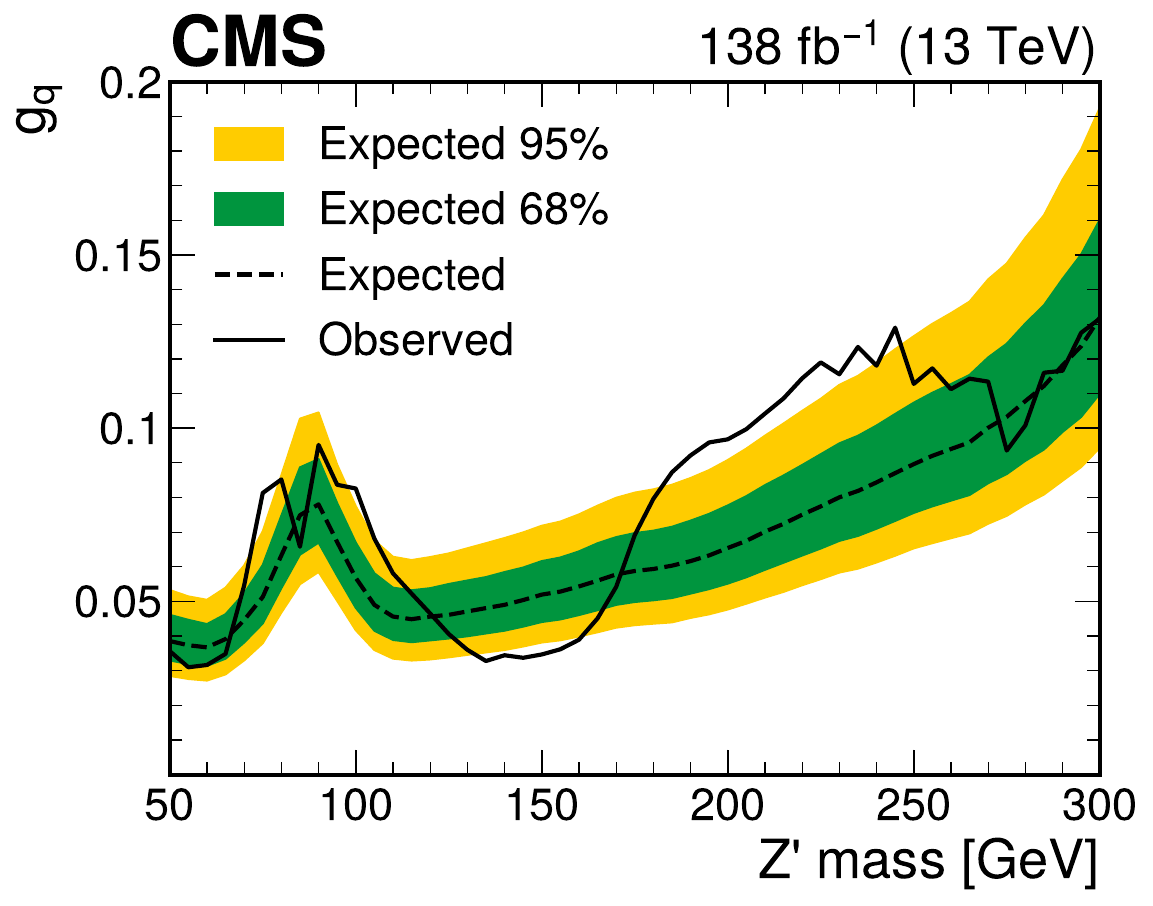}
    \includegraphics[width=0.48\textwidth]{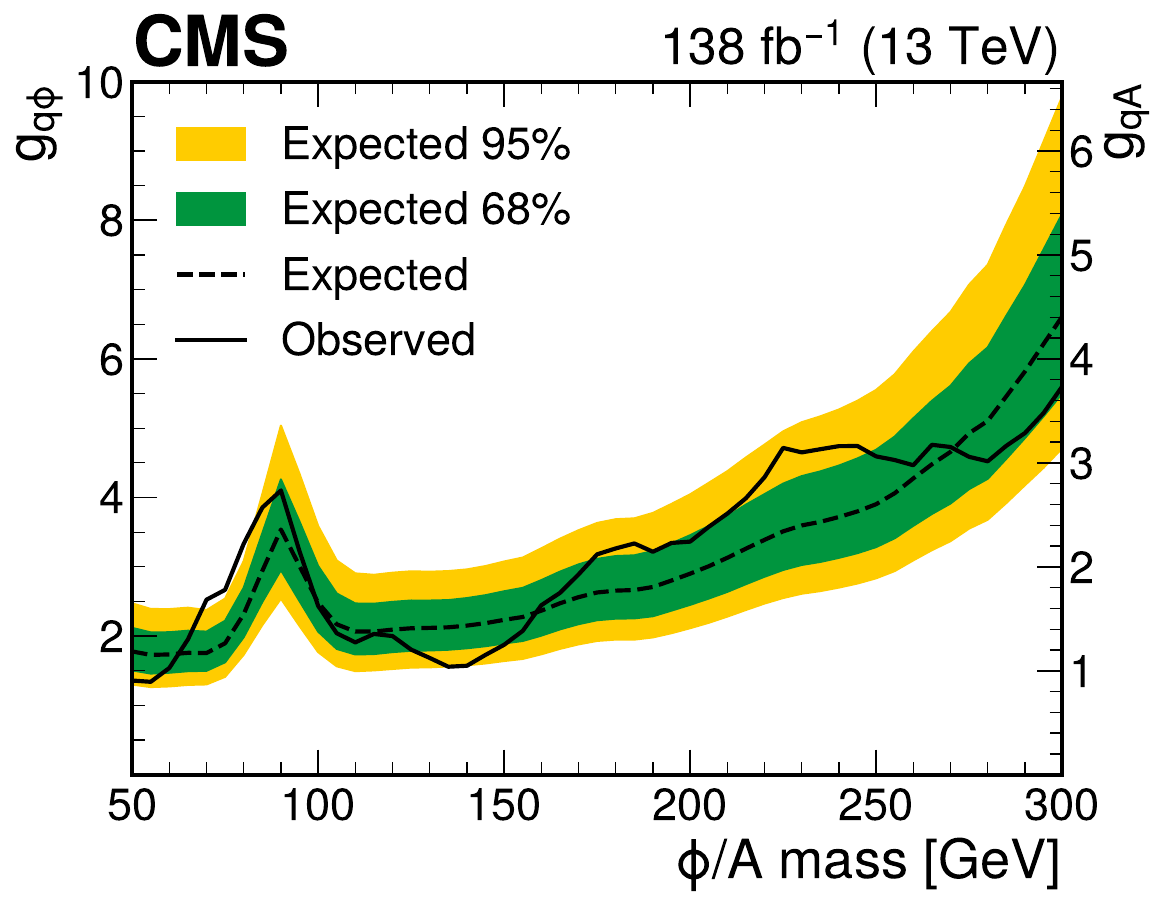}    
    \caption{
        Upper limits at 95\% \CL on the universal coupling \gq between the \PZpr boson and quarks (\cmsLeft) and the coupling parameters \gqphi (left vertical axis) and \gqA (right vertical axis) of the spin-0 model (\cmsRight). 
        The \PZpr boson is assumed to decay to quark-antiquark pairs with a 100\% branching fraction. 
        The $\phi$ and \PA bosons couple to quarks with a coupling given by \gqphi or \gqA times the SM Yukawa couplings, respectively, and decay dominantly to bottom quark-antiquark pairs. 
        The solid line represents the observed limits. The dashed line and the inner (yellow) and outer (blue) bands represent the median and the regions containing 68\% and 95\%, respectively, of the distribution of expected limits under the background-only hypothesis.
    }
    \label{fig:limits-spin1}
\end{figure}

Upper limits on the couplings of the resonances to quarks are set at 95\% confidence level (\CL) using a modified frequentist approach based on the \CLs criterion~\cite{CLS1,CLS2} with an asymptotic approximation for the test statistic~\cite{asymptotic}.
For the spin-1 model, the limits on the coupling \gq are shown in Fig.~\ref{fig:limits-spin1} (\cmsLeft), ranging from 0.03 to 0.13. 
Each signal category provides approximately the same sensitivity to \PZpr resonances.
The limits on \gqphi (\gqA) for the spin-0 model are shown in Fig.~\ref{fig:limits-spin1} (\cmsRight), and range from 1.5 to 5.8 (1.0 to 3.8), depending on the resonance mass.
For both the spin-1 and the spin-0 models, the limits are the most stringent to date on dijet resonances with masses between 50 and 250\GeV, improving upon the previous best limits by about a factor of two~\cite{CMS-boosted5,CMS-boosted3}. 
At higher resonance masses, the sensitivity decreases because the decay products are not fully contained by the AK8 jet algorithm; for resonance masses above 250\GeV, the most stringent limits are set by an ATLAS search targeting boosted signatures with two resolved jets~\cite{ATLAS:2024bms}.
Additional interpretations in the context of dark matter models are presented in \cmsEndMatter.

In summary, a search for new dijet resonances with masses from 50 to 300\GeV has been presented, using proton-proton collision data at $\sqrt{s}=13\TeV$, collected in 2016--2018 and corresponding to an integrated luminosity of 138\fbinv.
The search targets spin-1 resonances decaying to quark-antiquark pairs of all flavors, and spin-0 resonances decaying to bottom quark-antiquark pairs. 
The resonances are distinguished from the backgrounds using the \PNET algorithm and would result in narrow peaks in the jet soft-drop mass spectra. 
No significant excesses beyond the standard model expectations are observed. 
Limits are set on the couplings of the resonances to quarks, assuming that the resonances decay only to quarks. 
For spin-1 resonances coupled equally to all quark flavors, the upper limits on \gq range from 0.03 to 0.13.
For spin-0 resonances coupled dominantly to bottom quarks, the upper limits on \gqphi (\gqA) range from 1.5 to 5.8 (1.0 to 3.8).
These limits improve significantly on the existing ones in the mass range of 50--250\GeV.

\begin{acknowledgments}
We congratulate our colleagues in the CERN accelerator departments for the excellent performance of the LHC and thank the technical and administrative staffs at CERN and at other CMS institutes for their contributions to the success of the CMS effort. In addition, we gratefully acknowledge the computing centers and personnel of the Worldwide LHC Computing Grid and other centers for delivering so effectively the computing infrastructure essential to our analyses. Finally, we acknowledge the enduring support for the construction and operation of the LHC, the CMS detector, and the supporting computing infrastructure provided by the following funding agencies: SC (Armenia), BMBWF and FWF (Austria); FNRS and FWO (Belgium); CNPq, CAPES, FAPERJ, FAPERGS, and FAPESP (Brazil); MES and BNSF (Bulgaria); CERN; CAS, MoST, and NSFC (China); MINCIENCIAS (Colombia); MSES and CSF (Croatia); RIF (Cyprus); SENESCYT (Ecuador); ERC PRG and PSG, TARISTU24-TK10 and MoER TK202 (Estonia); Academy of Finland, MEC, and HIP (Finland); CEA and CNRS/IN2P3 (France); SRNSF (Georgia); BMFTR, DFG, and HGF (Germany); GSRI (Greece); NKFIH (Hungary); DAE and DST (India); IPM (Iran); SFI (Ireland); INFN (Italy); MSIT and NRF (Republic of Korea); MES (Latvia); LMTLT (Lithuania); MOE and UM (Malaysia); BUAP, CINVESTAV, CONACYT, LNS, SEP, and UASLP-FAI (Mexico); MOS (Montenegro); MBIE (New Zealand); PAEC (Pakistan); MES, NSC, and NAWA (Poland); FCT (Portugal); MESTD (Serbia); MICIU/AEI and PCTI (Spain); MOSTR (Sri Lanka); Swiss Funding Agencies (Switzerland); MST (Taipei); MHESI (Thailand); TUBITAK and TENMAK (T\"{u}rkiye); NASU (Ukraine); STFC (United Kingdom); DOE and NSF (USA).
\end{acknowledgments}\section*{Data availability} Release and preservation of data used by the CMS Collaboration as the basis for publications is guided by the  \href{https://doi.org/10.7483/OPENDATA.CMS.1BNU.8V1W}{CMS data preservation, re-use and open access policy}.

\bibliography{auto_generated}
\appendix

\section{Dark matter interpretations}\label{appendix}
A complementary approach to collider-based dark matter (\PDM) searches, especially in cases where the \PDM is too heavy to be produced directly, is to search for  a mediator (\ie, a portal), \PM, which connects the dark sector to the SM~\cite{CMS:2024zqs}. 
In a minimal scenario consisting of the mediator and the DM candidate, constraints from the observed relic density are driven directly by the DM-SM interactions~\cite{Albert:2017onk}. 
If $\mPDM < \mPM/2$, relic density constraints impose a strong bound on the maximum value of the ratio of the mediator-DM coupling to \mPM, above which the DM is overproduced.
If $\mPDM > \mPM$, a lower bound on the coupling cannot be easily computed, and the thermalization dynamics are more complicated. 
If $\mPM/2 < \mPDM < \mPM$, the relic density constraints impose a minimum value for the ratio of the DM-mediator coupling to \mPDM, which serves as an illustrative bound for the sensitivity of this analysis. 
The dijet bounds from this analysis are recast in terms of two DM benchmark models: a spin-1 dark photon (\PAprime) and a pseudoscalar, axion-like particle (ALP)~\cite{Essig:2013lka}. 

\begin{figure}[bpht!]
    \centering
    \includegraphics[width=0.49\textwidth]{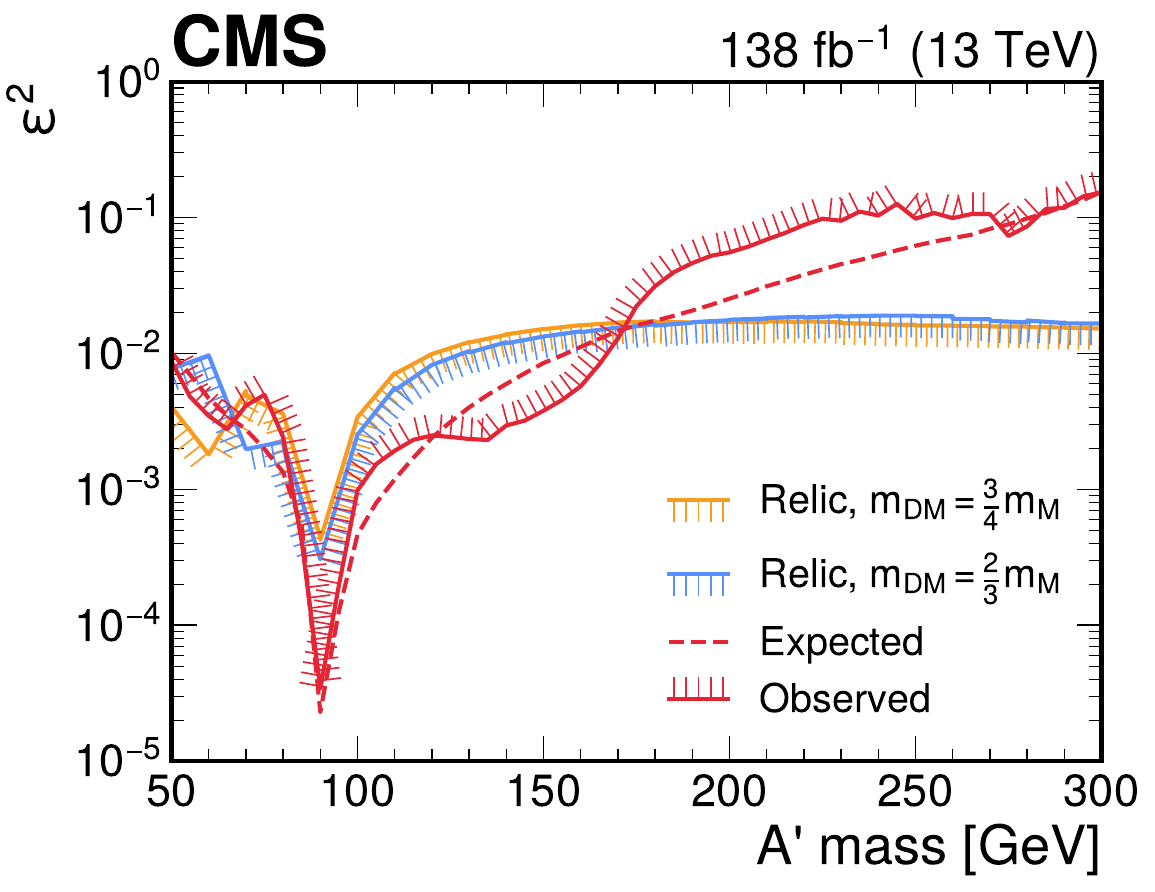}
    \includegraphics[width=0.49\textwidth]{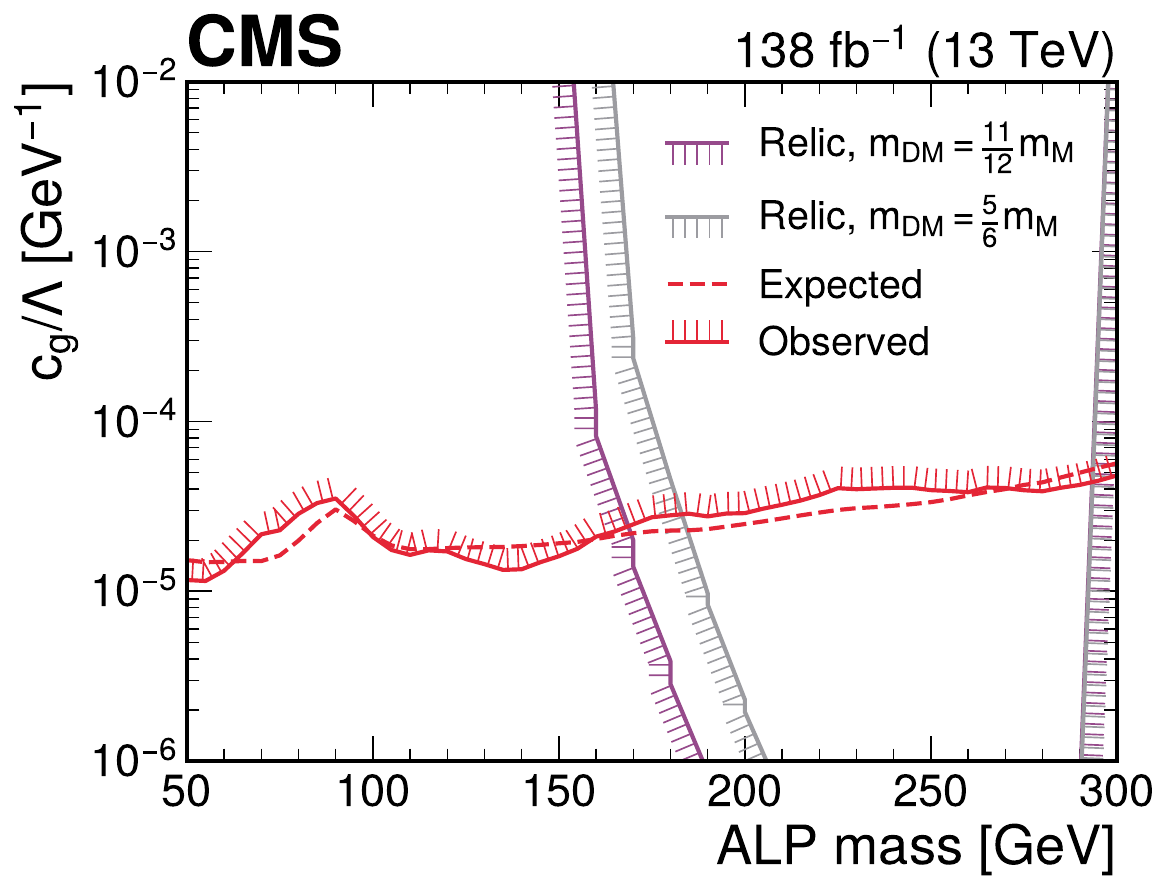}
    \caption{ 
     Recast of the 95\% \CL exclusion limits into the kinetic mixing parameter, $\epsilon^2$, of a dark-photon model (\PAprime, \cmsLeft) and the effective gluon coupling, $c_{\Pg}/\Lambda$, of an axion-like particle model (ALP, \cmsRight). 
     The solid (dashed) red lines represent the observed (expected) upper limits from this analysis.
     The orange and blue (\cmsLeft), and magenta and gray (\cmsRight) lines represent the minimum allowed coupling compatible with the observed DM relic density for representative values of $\mPDM/\mPM$.
     The hatched lines show the direction of the excluded regions.
    }
    \label{fig:dpbounds}
\end{figure}

For the simplest dark-photon model, the mixing parameter, $\epsilon$, can be approximated in terms of \gq to ${\approx} 20\%$ accuracy as:
\begin{equation*}
  \label{eq:eps_gq}
  \gq \simeq \epsilon e \frac{1}{\cos^{2}\thetaW} \frac{1}{1-\Delta_{\PZ}}\left(\sqrt{\langle Q^2\rangle}\cos^{2}\thetaW+\Delta_{\PZ}\sqrt{\langle Y^2\rangle}\right),
\end{equation*}
where \thetaW is the weak mixing angle, $\Delta_{\PZ}=(\mPAprime/\mPZ)^2$, and $\langle Q^2 \rangle \approx 1/3$ and $\langle Y^2 \rangle \approx 0.7$ are the average squared charge and hypercharge of incident partons in $\Pp\Pp$ collisions, respectively~\cite{CMS:2024zqs}.
Figure~\ref{fig:dpbounds} (\cmsLeft) shows the bounds on $\epsilon^2$. 
For $\mPAprime\lesssim 170\GeV$, the bounds exclude the minimal coupling compatible with the relic density for scenarios where $\frac{2}{3} \mPAprime < \mPDM < \mPAprime$. 

For the ALP model, the coupling to SM fermions is proportional to the fermion mass, and, hence, the ALP predominantly decays to bottom quarks. 
The limits on the pseudoscalar-quark coupling parameter, \gqA, are recast in terms of an effective gluon coupling, $c_{\Pg}/\Lambda$, assuming production via a top quark loop.
The limits are shown in Fig.~\ref{fig:dpbounds} (\cmsRight), and couplings greater than $c_{\Pg}/\Lambda\gtrsim(1\text{--}5)\times10^{-5}$ are excluded, depending on \mPALP.
For models with DM and an ALP mediator satisfying $(5/6) \mPALP \lesssim \mPDM < \mPALP$, a viable DM model is possible with the allowed coupling values and $200\lesssim \mPALP \lesssim 300\GeV$, but is otherwise excluded. 

\begin{figure}[htbp!]
    \centering
    \includegraphics[width=0.49\textwidth]{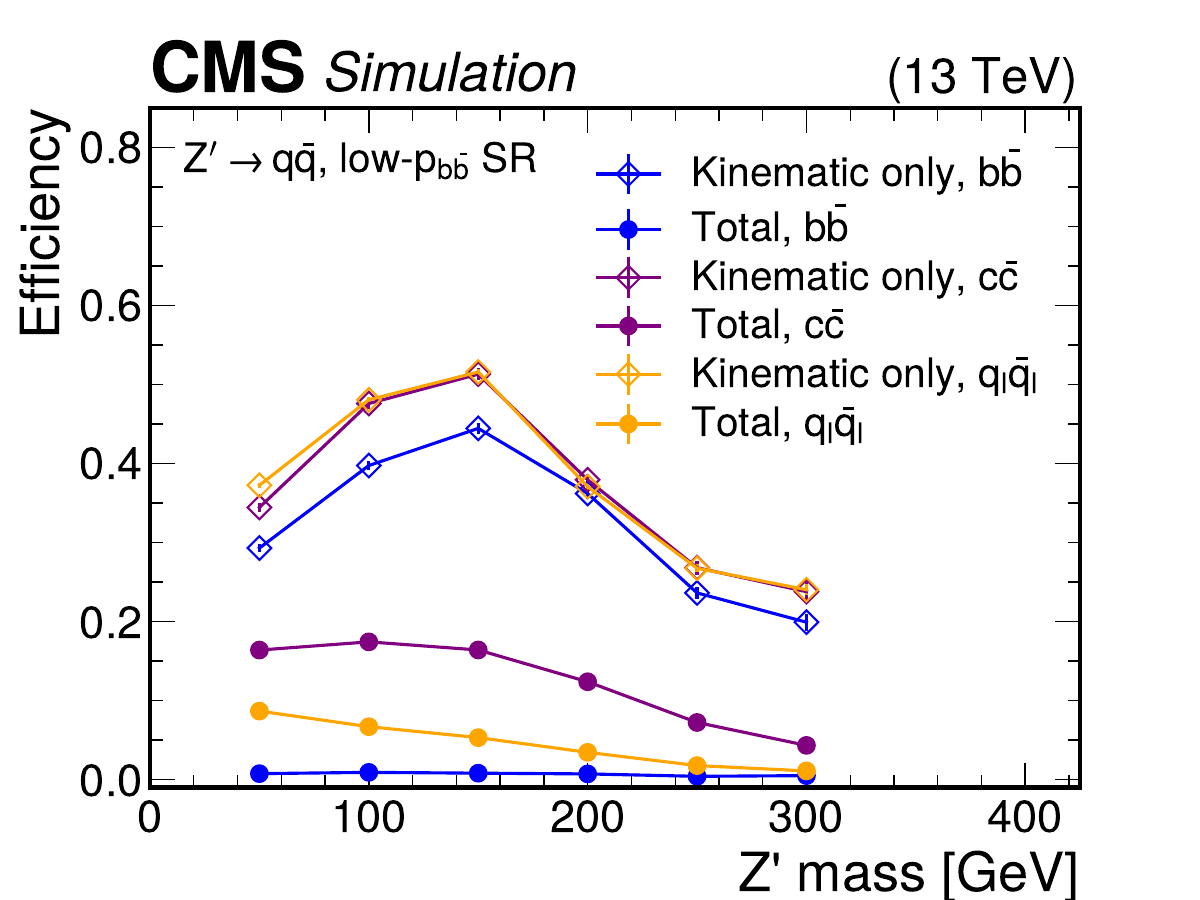}
    \includegraphics[width=0.49\textwidth]{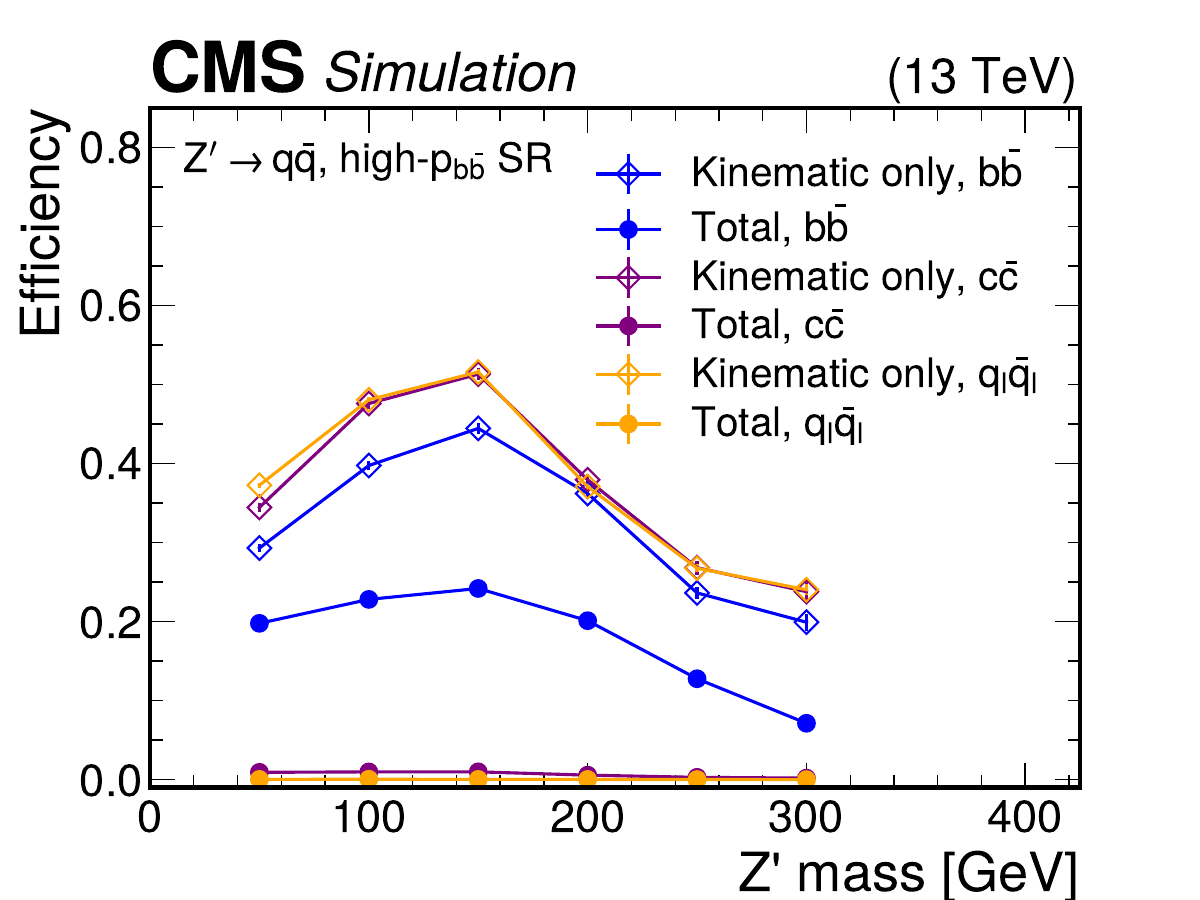}
    \caption{
        Efficiencies of the selection criteria in the low-\pbvcl (\cmsLeft) and high-\pbvcl (\cmsRight) signal regions, derived from simulated \PZpr events where the \PZpr satisfies $\pt>500\GeV$ and $\abseta<2.5$. 
        The efficiencies are shown separately for different quark flavors (denoted ``total'').  
        The efficiencies for all selection criteria, except for the \PNET requirements (denoted ``kinematic only''), are also shown to illustrate the performance of the jet tagging algorithm. 
    }
    \label{fig:model-independent-eff}
\end{figure}

\begin{figure}[htbp!]
    \centering
    \includegraphics[width=0.49\textwidth]{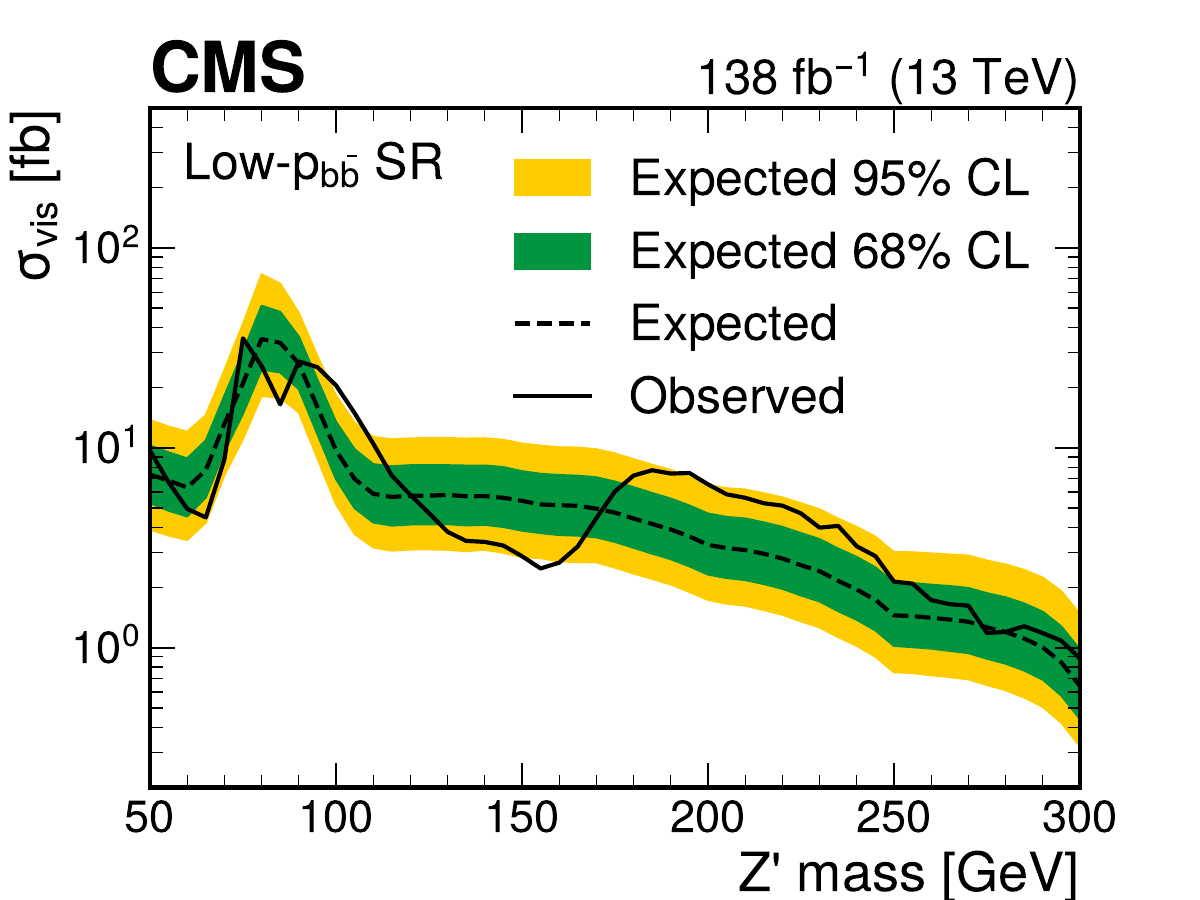}
    \includegraphics[width=0.49\textwidth]{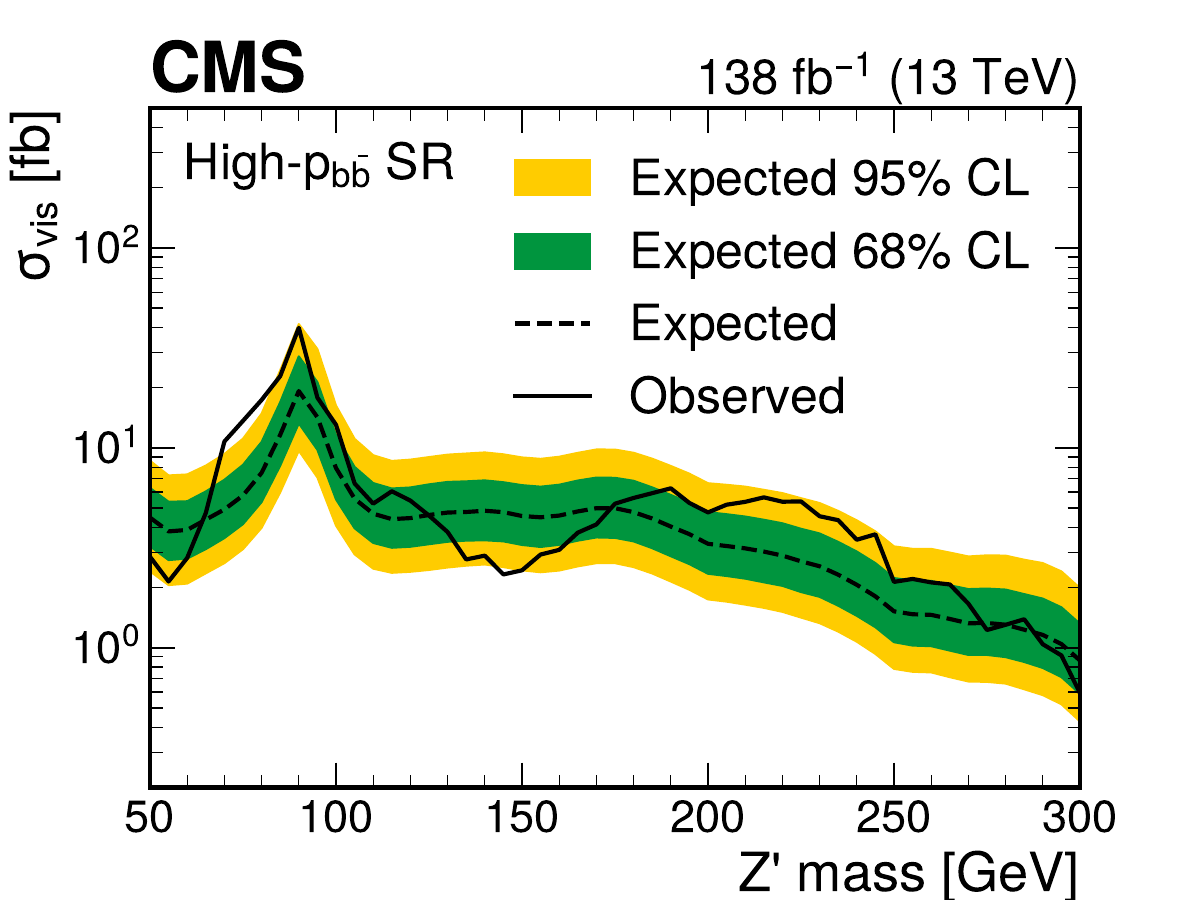}
    \caption{
        Upper limits at 95\% \CL on the visible cross section, $\sigma_{\text{vis}}$, defined as the number of signal events divided by the integrated luminosity, for the low-\pbvcl (\cmsLeft) and high-\pbvcl (\cmsRight) signal regions. 
        The solid line represents the observed limits, and the dashed line and the inner (green) and outer (yellow) bands represent the median and the regions containing 68\% and 95\%, respectively, of the distribution of expected limits under the background-only hypothesis.
    }
    \label{fig:model-independent-limits}
\end{figure}

To facilitate reinterpretation of the results, signal efficiencies and model-independent limits on the visible cross section, $\sigma_{\text{vis}}$, defined as the number of signal events divided by the integrated luminosity, are presented for each SR. 
The efficiencies, shown in Fig.~\ref{fig:model-independent-eff}, are computed using the simulated \PZpr sample, requiring that the event contain an AK8 jet reconstructed from stable generator-level particles, excluding neutrinos, satisfying $\pt>500\GeV$ and $\abseta<2.5$. 
To illustrate the performance of the jet tagging algorithm, the efficiencies for the selection criteria excluding the \PNET requirements are also shown. 
The model-independent limits are derived using a modified version of the binned maximum likelihood fit, incorporating the same CRs but only one SR (either low-\pbvcl or high-\pbvcl). 
The signal templates are taken from the \PZpr model, including decays to all flavors of quarks. 
The limits are shown in Fig.~\ref{fig:model-independent-limits}. 

\cleardoublepage \section{The CMS Collaboration \label{app:collab}}\begin{sloppypar}\hyphenpenalty=5000\widowpenalty=500\clubpenalty=5000\cmsinstitute{Yerevan Physics Institute, Yerevan, Armenia}
{\tolerance=6000
A.~Hayrapetyan, V.~Makarenko\cmsorcid{0000-0002-8406-8605}, A.~Tumasyan\cmsAuthorMark{1}\cmsorcid{0009-0000-0684-6742}
\par}
\cmsinstitute{Institut f\"{u}r Hochenergiephysik, Vienna, Austria}
{\tolerance=6000
W.~Adam\cmsorcid{0000-0001-9099-4341}, J.W.~Andrejkovic, L.~Benato\cmsorcid{0000-0001-5135-7489}, T.~Bergauer\cmsorcid{0000-0002-5786-0293}, K.~Damanakis\cmsorcid{0000-0001-5389-2872}, M.~Dragicevic\cmsorcid{0000-0003-1967-6783}, C.~Giordano\cmsorcid{0000-0001-6317-2481}, P.S.~Hussain\cmsorcid{0000-0002-4825-5278}, M.~Jeitler\cmsAuthorMark{2}\cmsorcid{0000-0002-5141-9560}, N.~Krammer\cmsorcid{0000-0002-0548-0985}, A.~Li\cmsorcid{0000-0002-4547-116X}, D.~Liko\cmsorcid{0000-0002-3380-473X}, I.~Mikulec\cmsorcid{0000-0003-0385-2746}, J.~Schieck\cmsAuthorMark{2}\cmsorcid{0000-0002-1058-8093}, R.~Sch\"{o}fbeck\cmsAuthorMark{2}\cmsorcid{0000-0002-2332-8784}, D.~Schwarz\cmsorcid{0000-0002-3821-7331}, M.~Shooshtari\cmsorcid{0009-0004-8882-4887}, M.~Sonawane\cmsorcid{0000-0003-0510-7010}, W.~Waltenberger\cmsorcid{0000-0002-6215-7228}, C.-E.~Wulz\cmsAuthorMark{2}\cmsorcid{0000-0001-9226-5812}
% ...
\par}
\vskip\cmsinstskip
\dag:~Deceased\\
$^{1}$Also at Yerevan State University, Yerevan, Armenia\\
$^{2}$Also at TU Wien, Vienna, Austria\\
% ...
\end{sloppypar}
\end{document}